\documentclass{elsart}
\usepackage[T1]{fontenc}
\usepackage{amssymb}
\usepackage{latexsym}
\usepackage{epsfig}

\newcommand{\1}{|1\rangle}
\newcommand{\0}{|0\rangle}
\newcommand{\tr}{\operatorname{tr}}

\begin{document}

\begin{frontmatter}
\title{A new entanglement measure induced by the Hilbert-Schmidt norm}

\author{C.~Witte},
\author{M.~Trucks} 
\address{Institut f\"ur Theoretische Physik\\
Technische Universit\"at Berlin\\
Hardenbergstra\ss e 36, 10623 Berlin, Germany\\
E-mail: christo@physik.tu-berlin.de, trucks@physik.tu-berlin.de}
\date{\today}
\maketitle
\begin{abstract}
  In this letter we discuss a new entanglement measure. It is based on
  the Hilbert-Schmidt norm of operators. We give an explicit formula
  for calculating the entanglement of a large set of states on
  $\mathbb{C}^{2}\otimes \mathbb{C}^2$. Furthermore we find some relations
  between the entanglement of relative entropy and the Hilbert-Schmidt
  entanglement. A rigorous definition of partial transposition is
  given in the appendix.
\end{abstract}
\end{frontmatter}

\section{Introduction}
\label{sec:intro}

Quantum information processing has received a considerable interest in
the last years, induced by the possibility of teleporting an unknown quantum
state and building a quantum computer. Also new
questions on the relation of quantum and classical physics arise in
this context. The feature which makes quantum computation more
efficient than classical computation and allows teleportation is
entanglement. Therefore there is also an increasing interest in
quantifying entanglement \cite{VPRK97}. Our letter considers the
quantification by introducing a new entanglement measure.

For pure states on the tensor product of two Hilbert spaces a measure
is given by the entanglement of entropy. Let ${\mathcal T}$ be the set
of states on the tensor product of two Hilbert spaces ${\mathcal
  H}_{1}\otimes{\mathcal H}_{2}$, i.e.~the set of all positive trace
class operators with trace 1. For a pure state $\sigma\in{\mathcal
  T}$, the entanglement of entropy $E(\sigma)$ is given by
\begin{eqnarray}\label{eq:pure-entanglement}
  E(\sigma) := - \mbox{tr}\;(\sigma_{1}\log_{2}\sigma_{1}) =
  -\mbox{tr}\;(\sigma_{2}\log_{2}\sigma_{2}) =
  -\sum_{i}|\alpha_{i}|^{2}\log_{2}|\alpha_{i}|^{2} 
\end{eqnarray}
where $\sigma_{i} = \mbox{tr}_{i}\;\sigma, i=1,2,$ are the partial
traces taken in the Hilbert spaces ${\mathcal H}_{i}$ and $\alpha_{i}$
are the Schmidt coefficients of $\sigma$ (cp.~\cite{Per}).

Considering mixed states, the situation is more complicated.  Several
entanglement measures have been defined in this case, e.g.\ the
entanglement of creation \cite{BBPS96} and the entanglement of
distillation \cite{BBPS96}. Here we follow an idea of Vedral et
al.~\cite{VPRK97}, based on measuring the distance between states in
the quantum mechanical state space. The set of disentangled states
${\mathcal D}$ is usually considered as the set of all states which
can be written as convex combinations of pure tensor states:
\begin{eqnarray*}
  {\mathcal D} := \{\rho\in{\mathcal T}\;|\; \rho = \sum_{i}
  p_{i}\rho^{(1)}_{i} \otimes \rho^{(2)}_{i}, \sum_{i}p_{i} = 1,
  \rho^{(k)}_{i}\in{\mathcal T}({\mathcal H}_{i}), k=1,2\}.
\end{eqnarray*}
The general idea of Vedral et al.~\cite{VPRK97} to quantify the amount
of entanglement of a state $\sigma \in {\mathcal T}\setminus{\mathcal
  D}$ is to define a distance of $\sigma$ to the set ${\mathcal D}$,
so that the entanglement $E$ of $\sigma$ is given by
\begin{eqnarray}\label{eq:entang}
  E(\sigma) = \min_{\rho\in{\mathcal D}} D(\sigma||\rho).
\end{eqnarray}
Here $D$ is any measure of distance between the density matrices
$\rho$ and $\sigma$, not necessarily a distance in the metrical
sense. There are several possibilities to define such a distance. One
example is the relative entropy $S(\sigma||\rho)$, given by
\begin{eqnarray*}
  S(\sigma||\rho) := \mbox{tr}\;(\sigma\log_{2}\sigma-\sigma\log_{2}\rho),
\end{eqnarray*}
discussed in \cite{VPRK97,VP98}. Another example is to take the Bures
metric as distance \cite{VP98}.

As a measure of distance we discuss in this letter the Hilbert-Schmidt
norm. The Hilbert-Schmidt norm is defined by
\begin{eqnarray*}
  \|A\|_{\mathrm{HS}}^{2} := \mbox{tr}\;(A^{*}A), 
\end{eqnarray*}
for all Hilbert-Schmidt operators on $\mathcal{H}={\mathcal
  H}_{1}\otimes{\mathcal H}_{2}$, i.e.\ for all operators if
$\dim\mathcal{H}<\infty$. We therefore define the Hilbert-Schmidt
entanglement (HS entanglement) $E_{\mathrm{HS}}$ of a state $\sigma$
by
\begin{eqnarray*}
  E_{\mathrm{HS}}(\sigma) := \min_{\rho\in{\mathcal
      D}}\|\rho-\sigma\|^{2}_{\mathrm{HS}}. 
\end{eqnarray*}
The choice of the squared distance instead of
$\|\rho-\sigma\|_{\mathrm{HS}}$ is motivated by the fact that it is
easier in calculations and justified because they are equivalent to
each other.

There are several requirements every measure of entanglement $E$
should satisfy (see e.g.~\cite{VPRK97,VP98} for a more detailed
discussion):
\begin{enumerate}
\item $E(\sigma) = 0$ for all $\sigma\in{\mathcal D}$.
\item \label{second_cond} $E(\sigma) = E(U_{1}\otimes U_{2}\sigma
  U_{1}^{*}\otimes U_{2}^{*})$ for all unitary operators $U_{i}\in
  {\mathcal H}_{i}, i=1,2,$ i.e.~the measure is invariant under local
  unitary operations.
\item The measure $E$ does not increase under local general
  measurements and classical communication, i.e.~for every completely
  positive trace-preserving map $\Theta : {\mathcal T} \to {\mathcal
    T}$ we have $E(\Theta\sigma) \leq E(\sigma)$.
\end{enumerate}
Of course every measure of entanglement defined by
eq.~(\ref{eq:entang}) trivially fulfils the first requirement.  It
can be seen as follows that condition~(\ref{second_cond}) is
satisfied: With $U = U_{1} \otimes U_{2}$, we have
\begin{eqnarray*}
  E(U \sigma U^*) &=&
  \min_{\rho\in{\mathcal D}} \|\rho - U\sigma U^*\|^{2} \\
  &=& \min_{\rho\in{\mathcal D}} \mbox{tr}\;(\rho^{2} - 2 \rho U \sigma
  U^{*} + U \sigma^{2} U^{*})\\
  &=& \min_{\tilde{\rho}\in{\mathcal D}} \mbox{tr}\;(\tilde{\rho}^{2} -
  2\tilde{\rho}\sigma + \sigma^{2}) \\
  &=& \min_{\rho\in{\mathcal D}} \|\rho-\sigma\|^{2} = E(\sigma),
\end{eqnarray*}
where we set $\tilde{\rho} = U^{*} \rho \,U \in{\mathcal D}$.  To show
that the third condition is fulfilled we apply a theorem of Lindblad
\cite{Lin74}.
\begin{thm}
  Let $\Phi:B({\mathcal H}) \to B({\mathcal H})$ be a positive
  mapping. Then
  \begin{eqnarray*}
    \|\Phi\| \leq 1, \quad \mathrm{tr}\;(\Phi A) = \mathrm{tr}\;A,
    \quad\forall A \in {\mathcal T}({\mathcal H}) \\
    \Longleftrightarrow \quad f(\Phi A) \leq f(A), \quad \forall f
    \;\mathrm{convex}, \quad A \in {\mathcal T}({\mathcal H}).
  \end{eqnarray*}
\end{thm}
Now since $\sigma \to E(\sigma)$ is a convex function, we conclude
that $E(\Theta\sigma) \leq E(\sigma)$.

\section{The HS-entanglement of some special states}

The use of geometric distance in the real vector space of selfadjoint
matrices as a measure of entanglement gives us the possibility to \it
see \rm the point of minimal distance in ${\mathcal D}$ (here referred
to as \it basepoint\rm) for some important cases easily. Recall that
the distance of an arbitrary point outside a convex and compact set
${\mathcal C}$ to this set is the closest distance to any orthogonal
projection of the point onto the (nontrivial) faces of ${\mathcal C}$
(A face $\mathcal{C}$ of a convex set $K$ is a convex subset of $K$
such that $\phi = \lambda\phi_{1} + (1-\lambda)\phi_{2}$ for
$\phi\in\mathcal{C}$, $\phi_{1,2}\in K$ and $0<\lambda<1$ imply
$\phi_{1,2}\in\mathcal{C}$. A face consisting of one point is an
extremal point of $K$. The trivial faces are the set $K$ itself and
the empty set).  The first set of states to be investigated are,
traditionally, the so-called \it Bell-states \rm on ${\mathcal
  H}=\mathbb{C}^2 \otimes \mathbb{C}^2$. These are expected to be \it
maximally entangled \rm for reasonable measures of entanglement. This
proves to be true also in this case.

Let us denote the basis of Bell-vectors corresponding to the natural
basis $\{\0,\1\}$ of $\mathbb{C}^2$ by $\psi_1=1/\sqrt{2}(\0 \0 \!+\!\1
\1)$, $\psi_2=1/\sqrt{2}(\0 \0 \!-\!\1 \1)$, $\psi_3=1/\sqrt{2}(\0
\1\! +\!\1 \0)$ and $\psi_4=1/\sqrt{2}(\0 \1 \!-\!\1 \0)$.
Furthermore $\Psi_i$ denotes the one-dimensional projector on the
vector $\psi_i$ (Bell-state) and $\Phi_{ij}=1/2(\Psi_i+\Psi_j)$ the
equally weighted mixture of $\Psi_i$ and $\Psi_j$. For a given
Bell-state $\Psi_i$ a Werner-state \cite{Wer89} is given by
$W_{\psi_i,\epsilon}=1/4(1-\epsilon) \mathbf{1}+\epsilon \Psi_i$, with
$\epsilon \in [0,1]$.  We can now formulate the following proposition,
which gives the entanglement of the Bell-states and arbitrary mixtures
of orthogonal Bell-states:
\begin{prop}
  For an arbitrary mixture of orthogonal Bell-states
  \begin{eqnarray*}
    \sigma=\lambda_i \Psi_i+\sum_{j\neq i} \lambda_j \Phi_{ij},
  \end{eqnarray*}
  where $\lambda_i \geq 0$ and $\sum \lambda_i =1$, the basepoint in
  ${\mathcal D}$ is given by
  \begin{eqnarray*}
    \tilde{\sigma}=\lambda_i W_{\psi_i,1/3}+\sum_{j\neq i} \lambda_j
    \Phi_{ij}
  \end{eqnarray*}
  and we find:
  \begin{eqnarray*}
    E_{\mathrm{HS}}(\sigma)=\lambda_i^2/3.
  \end{eqnarray*}
\end{prop}
Before we prove the proposition we give some remarks. Obviously, for a
given index $i$ we have found the entanglement of the Bell-state
$\Psi_i$ and all the states in the tetrahedron spanned by this
Bell-state and the three mixtures $\Phi_{ij}$. The complement of these
four tetrahedra in the larger tetrahedron of \it all \rm mixtures of
the four Bell-states is just the octahedron spanned by the six
disentangled states $\Phi_{ij}$ for $i \neq j$, which is therefore a
subset of the set ${\mathcal D}$. The fact that the states $\Phi_{ij}$
are in fact disentangled can be seen easily by either decomposition of
$\Phi_{ij}$ into disentangled projectors or partial transposition.
Thus all the mixtures of the four given Bell-states are covered by the
proposition.
\begin{pf*}{Proof of the proposition}
  We prove the fact that the suggested basepoint $\tilde{\sigma}$ is
  correct, by showing that the derivative of the function
  $f(\rho)=||\sigma-\rho||^2$ is non-negative at $\tilde{\sigma}$ in
  any direction leading \it into \rm the convex set ${\mathcal D}$.
  Such a directional derivative can be computed by using a
  parameterised line
  \begin{eqnarray*}
    \rho(s)=(1-s)\, \tilde{\sigma}+s \, \omega,
  \end{eqnarray*}
  where $\omega$ is an arbitrary element of ${\mathcal D}$ and
  calculating the derivative
  \begin{eqnarray}
    \frac{d}{ds}f(\rho(s))|_0&=&\lim_{s \to
      0}\frac{f(\rho(s))-f(\tilde{\sigma})}{s} \nonumber \\ 
    &=&\lim_{s \to 0}\frac{1}{s}
    \tr\bigl((\rho(s)-\sigma)^2-(\tilde{\sigma}-\sigma)^2\bigr)\nonumber\\ 
    &=&\lim_{s \to 0}\frac{1}{s} \tr\bigl(2 s (\sigma - \tilde{\sigma})
    (\tilde{\sigma} - \omega)+ s^2 (\tilde{\sigma}-\omega)^2\bigr)\nonumber\\
    &=&\tr\bigl(2 (\sigma - \tilde{\sigma}) (\tilde{\sigma} -
    \omega)\bigr).  \nonumber 
  \end{eqnarray} 
  We see that this derivative is an affine functional of the element
  $\omega \in {\mathcal D}$. Therefore convex combinations of elements
  in ${\mathcal D}$ lead to a convex combination of the result. For that
  reason it suffices to show non-negativity of the above expression only
  for $\omega \in \partial_e {\mathcal D}$, where $\partial_{e}$ denotes
  extreme points, in other words for disentangled projectors.
  
  Inserting the given expressions for $\sigma$ and $\tilde{\sigma}$ as
  well as choosing $\omega=P_\chi \otimes P_\xi$ to be a projector onto
  the normalised vectors $\chi$ and $\xi$, we get:
  \begin{eqnarray}
    \frac{d}{ds} f(\rho(s))|_0 &=& 2 \lambda_i \tr \bigl( (\Psi_i -
    W_{\psi_i,1/3})(\lambda_i W_{\psi_i,1/3}+\sum_{j \neq i}
    \lambda_j \Phi_{ij} - P_\chi \otimes P_\xi) \bigr)\nonumber  \\
    &=&2 \lambda_i \bigl(\lambda_i/2 + \sum_{j \neq
      i} \lambda_j/2 -\tr (\Psi_i P_\chi \otimes P_\xi) -
    \lambda_i/3 -\sum_{j \neq i}
    \lambda_j/3 \nonumber\\
    &&{}+\lambda_i/6+ \tr (\Psi_i P_\chi
    \otimes P_\xi)/3 \bigr)\nonumber  \\ 
    &=&2\lambda_i (1- 2 \tr (\Psi_i P_\chi \otimes P_\xi ))/3
    \nonumber  \\ 
    &=&2 \lambda_i (1- 2 \langle \psi_i,\chi \otimes \xi
    \rangle \langle \chi \otimes \xi, \psi_i \rangle  )/3 \nonumber \\ 
    &=&2 \lambda_i (1- 2 |\langle \chi \otimes
    \xi,\psi_i\rangle |^2)/3. \nonumber 
  \end{eqnarray} 
  Since any $\psi_i$ is of the form $ (\0\otimes U\0 +
  \1\otimes U \1)/\sqrt{2}$, where $U$ is a unitary transformation, we can
  write:
  \begin{eqnarray}
    \frac{d}{ds}f(\rho(s))|_0&=& 2 \lambda_i (1- |\langle
    \chi \0 \langle U^* \xi\0+
    \langle \chi \1 \langle U^* \xi\1|^2)/3 \nonumber\\
    &=& 2 \lambda_i (1- | \chi_0 (U^*\xi)_0+\chi_1
    (U^*\xi)_1|^2)/3 \nonumber \\ 
    &=& 2 \lambda_i (1- | \langle
    \bar{\chi},U^*\xi\rangle|^2)/3     \nonumber  \\ 
    &\geq& 0.\nonumber 
  \end{eqnarray}
  The final step is just an application of the Cauchy-Schwartz
  inequality.  \hfill \qed
\end{pf*}
The next class of states we are going to deal with also includes
the Bell-states as special case, namely the \it pure states. \rm
Unfortunately, the geometry of the underlying part of the face of
${\mathcal D}$ proves to be somewhat more complex. This leads to the
fact that pure states admit an \it easy-to-construct \rm basepoint
only under a certain condition, which is stated in the following
proposition:
\begin{prop} \label{pro:pure}
  Let $\phi$ be a vector in $\mathbb{C}^2\otimes\mathbb{C}^2$, written in its
  Schmidt-basis as $\phi=a\0\otimes\0+b\1 \otimes\1$ with $a$ and $b$
  positive numbers, such that
  \begin{eqnarray*}
    a^2 \in\bigl[\frac{1}{2}-\frac{\sqrt{5}}{6},\frac{1}{2} + 
    \frac{\sqrt{5}}{6}\bigr] \quad;\quad b^2=1-a^2.
  \end{eqnarray*}
  The basepoint associated to the one-dimensional entangled projector
  $\sigma$ onto the span of $\phi$ is then given by:
  \begin{eqnarray*}
    \tilde{\sigma}=\sigma-a b(4 P_{\psi_1}-\mathbf{1})/3,
  \end{eqnarray*}
  where $P_{\psi_1}$ is the Bell-state for $\psi_1=\0\0+\1\1$.
  The entanglement of $\sigma$ is given by:
  \begin{eqnarray*}
    E_{\mathrm{HS}}(\sigma)=4 a^2 b^2/3.
  \end{eqnarray*}
\end{prop}
\begin{pf}
  The suggested basepoint $\tilde{\sigma}$ has to be shown to lie in
  ${\mathcal{D}}$, first.  According to the \it Peres criterion \rm
  \cite{HHH96c} it suffices to show that $\tilde{\sigma}$ as well as
  its partial transpose are positive. A rigorous definition of \it
  partial transposition \rm is given in the appendix.  In this case
  $\sigma^{T_2}$ (partial transposition of $\sigma$ in the second
  factor) has exactly one negative eigenvalue, which can be seen by
  writing:
  \begin{eqnarray*}
    \sigma=a^2 P_{\0\0}+b^2 P_{\1\1}+a b P_+^{T_2}-a b P_-^{T_2},
  \end{eqnarray*} 
  where $P_+$ denotes the projector onto $\mbox{span}(\0\1+\1\0)$ and
  $P_-$ denotes the projector onto $\mbox{span}(\0\1-\1\0)$. 
  $\tilde{\sigma}$ was found by projecting onto the
  corresponding plane with eigenvalue zero ($\{\rho:\tr \rho=1 \wedge
  \tr (P_-
^{T_2} \rho)=0\}$), thus:
  \begin{eqnarray*}
    \tilde{\sigma}=(a^2-a b/3) P_{\0\0}+(b^2-a b/3) P_{\1\1}+2 a b 
    P_{+}^{T_2}/3.
  \end{eqnarray*}
  $\tilde{\sigma}^{T_2}$ is positive, iff $a^2-a
  b/3>0$ and $b^2-ab/3>0$. This yields the condition:
  \begin{eqnarray*}
    a^2 \in [1/10,9/10].
  \end{eqnarray*}
  The stronger condition is nevertheless the positivity of
  $\tilde{\sigma}$ itself. We find that the only possibly negative
  eigenvalue has to satisfy the following inequality:
  \begin{eqnarray}
    3-2 a b-\sqrt{9 a^4-14 a^2 b^2+9 b^4}&\geq&0 \\
    (3-2 a b)^2&\geq& 9 a^4- 14 a^2 b^2 +9 b^4 \nonumber\\
    12 a b &\geq&36 (a^2-a^4)\nonumber\\
    1 &\geq& 3 a b. \nonumber 
  \end{eqnarray}
  This inequality is exactly satisfied for $a^2$ in the stated interval.

  It remains to prove minimality of the distance of the basepoint to the
  pure state $\sigma$.  In the notation of the preceding proof we have
  to calculate here:
  \begin{eqnarray}
    \frac{d}{ds}f(\rho(s))|_0&=&\tr\bigl(2 (\sigma - \tilde{\sigma})
    (\tilde{\sigma} - \omega)\bigr) \nonumber 
  \end{eqnarray} 
  and show that this expression is non-negative for any disentangled
  one-dimensional projector $\omega=P_\xi \otimes P\chi$.
  
  We find:
  \begin{eqnarray}
    \tr(\sigma-\tilde{\sigma})\tilde{\sigma}&=&a b
    \tr((4P_{\psi_1}-\mathbf{1})(\tilde{\sigma}))/3\nonumber \\ 
    &=&a b  (\tr(4P_{\psi_1}\tilde{\sigma})- 1)/3 \nonumber \\
    &=&a b (-4 a b +4 \tr (P_{\psi_1} \sigma)- 1)/3\nonumber \\
    &=&a b /3 \nonumber 
  \end{eqnarray} 
  Finally:
  \begin{eqnarray}
    \frac{d}{ds}f(\rho(s))|_0&=& 2 a b [ 1 - \tr((4
    P_{\psi_1}-\mathbf{1})P_\xi\otimes P_\chi))]/3 \nonumber  \\ 
    &=& 2 a b (2- 4\tr(P_{\psi_1} P_\xi\otimes P_\chi))/3,\nonumber
  \end{eqnarray} 
  which is, except for the leading positive factor, the same expression
  as in the preceding proof and thus non-negative.
  
  The explicit quantity of the entanglement is a pure matter of
  calculation.  \hfill \qed
\end{pf}


\begin{figure}[t]
  \begin{center}
    \includegraphics{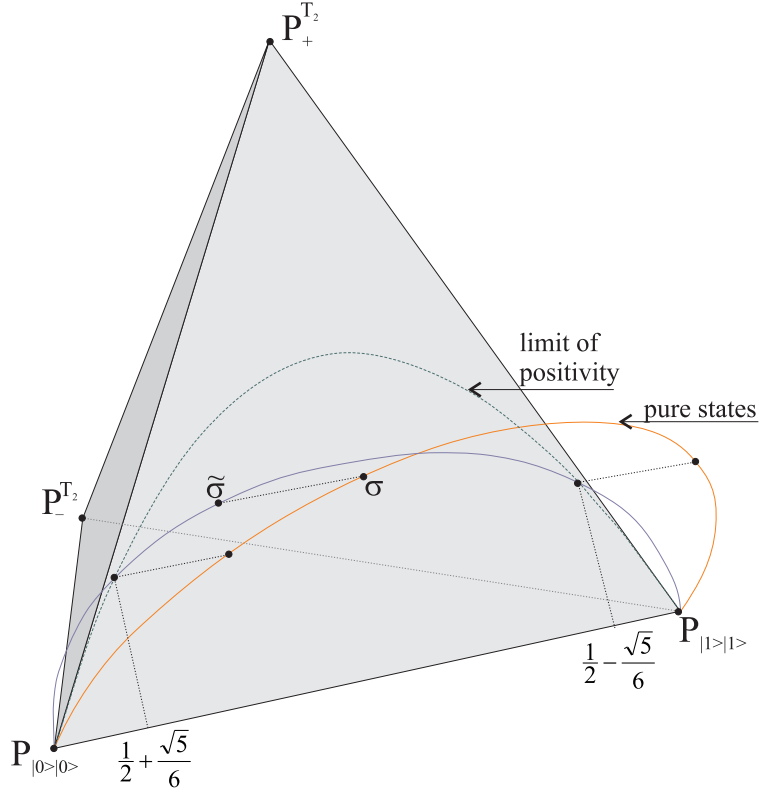}  
  \end{center}
  \caption{Pure states and their projection: Pure states lie on the
    marked semicircle and are projected onto the triangle spanned by
    the three partially transposed projectors in the front. The line
    formed by the projected semicircle is a segment of an ellipse that
    intersects the parabola marked as limit of positivity at the two
    critical Schmidt coefficients. Only the states below this parabola
    are positive. The projected states above the limit of positivity
    are therefore not admissible.}
\end{figure}
  
For the remaining pure states the calculation of the entanglement is a
bit more complicated. For this purpose we first parameterise
the parabola that forms the border of the positive elements in the
triangle $conv (\{ P_{\0\0},P_{\1\1},P_+^{T_2}\})$ (see fig.\ 1):
\begin{eqnarray}
   p(s)&:=&s^2  P_{\0\0}+(1-2 s+s^2)P_{\1\1}+2(s-s^2)P_+^{T_2} \nonumber
\end{eqnarray}
It is easy to see that this is a parabola, indeed, and explicit
calculation of the eigenvalues shows that the elements have a zero
eigenvalue. The idea is now to project onto this parabola, instead of
the whole triangle. Finding the minimal distance of the given pure
state to the parabola corresponds to minimising the function
\begin{equation}
  f(s)=\tr(p(s)-P_\phi)^2
\end{equation}
leading to a third order equation in the parameter $s$. Even though
rather cumbersome, this case illustrates the problems in finding
explicit solutions for entanglement measures.  We state the solution
here as a conjecture only, and will give the proof elsewhere.
\begin{conj}
  Let $\phi$ be a vector in $\mathbb{C}^2\otimes\mathbb{C}^2$, written in its
  Schmidt-basis as $\phi=a\0\otimes\0+b\1 \otimes\1$ with $a$ and $b$
  positive numbers, such that
  \begin{eqnarray*}
    a^2 \in \bigl[0,\frac{1}{2}-\frac{\sqrt{5}}{6}\bigr] \cup 
    \bigl[\frac{1}{2}+\frac{\sqrt{5}}{6},1\bigr] \quad;\quad 
    b^2=1-a^2.
  \end{eqnarray*} 
  The basepoint associated to the one-dimensional entangled projector
  $\sigma$ onto the span of $\phi$ is then given by:
  \begin{equation}
    t^2  P_{\0\0}+(1-2 t+t^2)P_{\1\1}+2(t-t^2)P_+^{T_2}, 
  \end{equation}
  where $t \in \Rset$ is the real solution of the cubic equation
  \begin{equation}
    -1 - a b + b^2 + (5  - a^2  + 2 a b  - b^2 )t - 9 t^2 + 
    6 t^3=0.  
  \end{equation}
\end{conj}
Finally we state a corollary that is independent on the conjecture
above:
\begin{cor}
  The Bell-states, i.e.~the projectors associated to vectors of the
  form $U\0\0 \otimes V\1\1$, where $U$ and $V$ are unitary operators
  on the single particle Hilbert spaces, are maximally entangled w.r.t
  the HS-entanglement.
\end{cor}
\begin{pf} 
  Every pure state has a HS-entanglement less or equal to the
  Bell-states (which evaluates to $1/3$).  This is obvious for those
  states covered be Prop.~\ref{pro:pure}, as seen by evaluating the
  explicitly given formula for the entanglement. For those states
  covered by the conjecture, we find, even if the exact value of the
  entanglement is unknown, the following inequality:
  \begin{eqnarray}
    E(P_\phi)&\leq &\tr(P_\phi P_{\0\0})^2 \nonumber \\
    &=& 2-2 \tr(P_\phi P_{\0\0}) \nonumber \\
    &=&2-2 a^2 \\
    &\leq & 1-\sqrt{5}/3 \quad \quad (*)\nonumber\\
    &\leq & 1/3, \nonumber 
  \end{eqnarray}
  where $(*)$ is valid for those states with $a^2 \in
  [1/2+\sqrt{5}/6,1]$. For those states with $a^2 \in
  [0,1/2-\sqrt{5}/6]$ the same argument using $P_{\1\1}$ instead of
  $P_{\0\0}$ is valid.
  
  We conclude that mixed states have an entanglement that is less or
  equal to its most entangled spectral projector (decomposed to dimension
  one), because the HS-entanglement is a convex function and spectral
  decomposition of operators in $\mathcal{T}$ is a convex
  combination. \hfill \qed 
\end{pf}
\begin{rem}
  Obviously not only the pure states, but also the mixture of each of
  these and their associated basepoint are analysed by our method.
  Convex combinations of a given state and its basepoint share, of
  course, the same basepoint. Also, it is easy to see that their
  entanglement is given by
\begin{equation}
  E_{\mathrm{HS}}(\lambda \sigma+(1-\lambda)\tilde{\sigma})=\lambda^2
  E_{\mathrm{HS}}(\sigma). 
\end{equation}
\end{rem} 

\section{The use of HS-entanglement}

The most obvious use of the HS-entanglement is its easy form, which
makes an explicit calculation possible by merely knowing the geometric
structure of the set of disentangled states. On the other hand it also
has the more practical property of yielding useful estimates for other
measures of entanglement. As an example we give an inequality
connecting the HS-entanglement to the very useful measure based on the
relative entropy (referred to as $E_{\mathrm{vN}}$ here).
\begin{prop}
  For any entangled state $\sigma \in {\mathcal T}$ the following
  inequality holds:
  \begin{eqnarray*}
    E_{\mathrm{vN}}(\sigma)\geq\frac{1}{2 \log_2
      2}E_{\mathrm{HS}}(\sigma)
  \end{eqnarray*}   
\end{prop}
\begin{pf}
  We denote the basepoints of $\sigma$ in ${\mathcal D}$
  w.r.t.\  the relative entropy by $\hat{\sigma}$ and w.r.t.\  the
  distance by $\tilde{\sigma}$. We get:
  \begin{eqnarray*}  
    E_{\mathrm{vN}}(\sigma) &=& S(\sigma || \hat{\sigma})\nonumber\\
    &\geq& \frac{1}{2\log_2 2}||\sigma-\hat{\sigma}||^2 \nonumber\\
    &\geq& \frac{1}{2 \log_2 2} || \sigma - \tilde{\sigma}||^2
    \nonumber \\ &=& \frac{1}{2\log_2 2}E_{\mathrm{HS}}(\sigma),
  \end{eqnarray*} 
  where we used a well-known estimate for the relative entropy (cf.
  \cite[Prop.1.1]{OP93}), adjusted to the use of $\log_2$ instead of
  $\ln$.  \hfill \qed
\end{pf}
\begin{rem}
  For pure states on the tensor product of two Hilbert spaces the
  entanglement of relative entropy is given by
  \begin{eqnarray*}
    E_{\mathrm{vN}}(\sigma) = - a^{2}\log_{2}a^{2} -
    b^{2}\log_{2}b^{2} = -a^{2}\log_{2}a^{2} -
    (1-a^{2})\log_{2}(1-a^{2}) 
  \end{eqnarray*}
  where $a$ and $b$ are the Schmidt coefficients of $\psi$, $\sigma =
  |\psi \rangle \langle \psi|$ (see Eq.~(\ref{eq:pure-entanglement})).
  The Hilbert-Schmidt entanglement gives the same order as this
  entanglement measure on pure states on $\mathbb{C}^{2} \otimes
  \mathbb{C}^{2}$, i.e.
  \begin{equation}
    E_{\mathrm{vN}}(\sigma) \leq E_{\mathrm{vN}}(\rho) \iff
    E_{\mathrm{HS}}(\sigma) \leq E_{\mathrm{HS}}(\rho)
  \end{equation}
  for pure states $\rho$ and $\sigma$. E.g.\  we have shown in
  Prop.~\ref{pro:pure}, that
  \begin{eqnarray*}
    E_{\mathrm{HS}}(\sigma) = 4a^{2}b^{2}/3 = 4a^{2}(1-a^{2})/3.
  \end{eqnarray*}
  Considering $E_{\mathrm{vN}}$ and $E_{\mathrm{HS}}$ as functions of
  $a$, both functions attain their maximum at $1/\sqrt{2}$ and are
  strictly increasing resp.\ decreasing for $a<1/\sqrt{2}$ resp.\ 
  $a>1/\sqrt{2}$, therefore give the same order for pure states on
  $\mathbb{C}^{2} \otimes \mathbb{C}^{2}$.
\end{rem}

\appendix

\section{Partial Transposition}

In a matrix-algebra ${\mathcal M}_\mathbb{C}(n)$ the concept of
transposition is intuitively defined by the mapping
\begin{eqnarray*} 
  T:\left( \begin{array}{ccc} a_{11} & \cdots & a_{1n} \\
      \vdots & \ddots & \vdots \\
      a_{n1} & \cdots & a_{nn}  \end{array}\right)
  \mapsto \left( \begin{array}{ccc} a_{11} & \cdots & a_{n1} \\
      \vdots & \ddots & \vdots \\
      a_{1n} & \cdots & a_{nn}  \end{array}\right).
\end{eqnarray*}
Nevertheless the concept seems to be far from natural, if the algebra
is not given as a finite dimensional matrix-algebra, but as the
operator algebra ${\mathcal B}({\mathcal H})$ over an abstract Hilbert
space ${\mathcal H}$.  Even if the Hilbert space is finite
dimensional, the mapping above is only defined, if a basis is chosen,
and is depending on that choice.  It is a well known fact that only
the concept of the \it adjoint \rm operator is given by the algebraic
properties of a complex space. \it Transposition \rm is a concept
connected to real vector spaces.

A rigorous definition of transposition in the complex case is possible
if a further structure is given to the complex Hilbert space
${\mathcal H}$. Basically this structure can be thought of as a \it
split into a real and an imaginary part.\rm
\begin{defn}
  Given a real Hilbert space ${\mathcal R}$ and a real linear
  isomorphism $K:{\mathcal R}\oplus{\mathcal R} \to {\mathcal H}$,
  such that $[x,y]+i[I x,y]=\langle K(x), K(y)\rangle\quad \forall
  x,y$, where $[.,.]$ is the scalar product on ${\mathcal
    R}\oplus{\mathcal R}$, $I:{\mathcal R}\oplus{\mathcal R} \to
  {\mathcal R}\oplus{\mathcal R},(x_r,x_i)\mapsto (-x_i,x_r)$ the
  canonical complexification and $\langle.,.\rangle$ the scalar
  product on ${\mathcal H}$.  The \bf transposition \it in ${\mathcal
    B}({\mathcal H})$ with respect to $K$ is then defined by the
  following equation:
  \begin{eqnarray*}K^*(A^T):=(K^*A)^T\end{eqnarray*}
\end{defn}
Obviously any choice of a basis in ${\mathcal H}$ defines a split into
real and imaginary part. The definition of the transposed matrix above
agrees with the new one for ${\mathcal H}=\mathbb{C}^n$. Any transposition
is a complex linear, involutive, positive mapping ${\mathcal
  B}({\mathcal H})\to{\mathcal B}({\mathcal H})$.  The composition of
two transpositions $T$ and $\tilde{T}$ can always be written in the
form $T \circ \tilde{T}(A)=U A U^*$, where $U$ is unitary. Thus the
composition is completely positive.
\begin{defn}
  Given a transposition $T$ on ${\mathcal B}({\mathcal H}_2)$, the
  {\bf partial transposition} (in the second factor) $T_2$ on
  ${\mathcal B}({\mathcal H}_1)\otimes {\mathcal B}({\mathcal H}_2)$
  is defined by:
  \begin{eqnarray*}
    T_2:=Id \otimes T
  \end{eqnarray*}
\end{defn}
Except for the trivial case that ${\mathcal H}_1$ is one-dimensional
the partial transposition is never a positive mapping. Nevertheless
the composition of two partial transpositions is always positive due
to the complete positivity of the composition of two transpositions.
This has the important consequence that the set of partially transposed
positive operators on a product algebra is independent of the choice
of transposition (cp.~\cite{HHH96c}).
\begin{ack}
  We thank K.-E.~Hellwig and W.~Stulpe for helpful hints and
  discussions.
\end{ack}

\end{document}